\documentclass{elsart}
\usepackage{amssymb}
\usepackage{amsmath}
\usepackage{rotating}

\input{tcilatex}

\begin{document}

\begin{frontmatter}

\title{Utility function estimation: the entropy approach}

\author{Andreia Dionisio* and A. Heitor Reis**}

\address{*University of Evora, Center of Business Studies, CEFAGE-UE, Largo Colegiais, 2, 7000 Evora, Portugal, E-mail: andreia@uevora.pt
** University of Evora, Department of Physics and Evora
Geophysics Center, ahr@uevora.pt }

\begin{abstract}
The maximum entropy principle can be used to assign utility values when only partial information is available about the decision maker's preferences. In order to obtain such utility values it is necessary to establish an analogy between probability and utility through the notion of a utility density function. According to some authors [Soofi (1990), Abbas (2006a) (2006b), Sandow et al. (2006), Friedman and Sandow (2006), Darooneh (2006)] the maximum entropy utility solution embeds a large family of utility functions.
In this paper we explore the maximum entropy principle to estimate the utility function of a risk averse decision maker.
\end{abstract}


\end{frontmatter}

\section*{Introduction}

Utility function is one of the most useful concepts in decision analysis and
may be computed empirically from analysis of a trading data of an agent
which demonstrates its tolerance with respect to risk. This concept
characterizes the excess demand in analogy to the potential energy in
mechanical systems [Darooneh (2006)] and the maximization of utility shows
the equilibrium condition of the respective market. The randomness in the
market tends to increase with time, which is a consequence of the existent
risks. Given this, the state of the market with maximum randomness or
uncertainty is called equilibrium.

A possible approach to estimate utility functions and utility values using
only partial information about the agent's preferences is the \emph{Maximum
Entropy }(ME) principle. In this paper we refer to partial information when
we only have inferred the utility values based on observed decisions.

The main assumption to derive the utility function of an agent, using the ME
principle is the correspondence between the concept of equilibrium in
physics (statistical) and economics (mechanical). According to some authors
[namely Foley (1994), Candeal \emph{et al.} (2001), Darooneh (2006)] the
economic equilibrium can be viewed as an asymptotic approximation to
physical equilibrium and some difficulties with mechanical picture
(economic) of the equilibrium may be improved by considering the statistical
(physical) description of it.

In this paper we explore the ME principle to estimate the utility values of
a risk averse investor. The rest of the paper is organized as follows. In
Section 1 we present a brief discussion of the background theory, namely the
ME principle and its applications to economics and more specifically to
decision analysis. Section 2 presents the analogy between utility and
probability, and utility and entropy. Finally, Section 3 presents the\ main
conclusions of this study.

\section{Background theory}

Suppose that we have a set of possible events whose probabilities of
occurrence are $p_{1}$, $p_{2},...,p_{n}$ and $H$ is a measure of
uncertainty. Shannon (1948) developed a measure of uncertainty associated
with an outcome from a set of symbols that satisfy the following properties $%
\left( i\right) $ $H$ should be continuous in $p_{i},i=1,...,n;$ $\left(
ii\right) $\ if $p_{i}=1/n$, then $H$ should be a monotonic increasing
function of $n$; $\left( iii\right) \ H$ is maximized in a uniform
probability distribution context; $\left( iv\right) \ H$ should be additive; 
$\left( v\right) \ H$ should be the weighted sum of the individual values of 
$H.$

According to Shannon (1948) a measure that satisfies all these properties is
the entropy which is defined as $H\left( X\right) =-\tsum_{i}p_{i}\log p_{i}.
$When the random variable has a continuous distribution, and $p_{X}(x)$ is
the density function of the random variable $X$, the entropy (usually called
differential entropy) is given by $H\left( X\right) =-\int p_{X}(x)\log
p_{X}(x)dx.$

The properties of the entropy of continuous (differential entropy) and
discrete distributions are mainly alike. For continuous distributions, $%
H\left( X\right) $ is not scale invariant $\left( H\left( cX\right) =H\left(
X\right) +\log \left\vert c\right\vert \right) $ but is translations
invariant $\left( H\left( c+X\right) =H\left( X\right) \right) $. The
differential entropy may be negative and infinite [Shannon (1948), Soofi
(1994)]. Entropy $\left[ H\left( X\right) \right] $ is a measure of the
average amount of information provided by an outcome of a random variable
and similarly, is a measure of uncertainty about a specific possible outcome
before observing it [Golan (2002)].

Jaynes (1957) introduced the maximum entropy (ME) principle as a
generalization of Laplace's principle of insufficient reason. The ME
principle appears as the best way when we intend to make inference about an
unknown distribution based only on few restrictions conditions, which
represent some moments of the distribution. According to several authors
[see for example Soofi (2000) and Golan (2002)] this principle uses only
relevant information and eliminates all irrelevant details from the
calculations by averaging over them.

The ME model is usually formulated to confirm the equality constraints on
moments or cumulative probabilities of the distribution of the random
variable $X,$ where $h_{j}\left( X_{i}\right) $ is an indicator function
over an interval for cumulative probability constraints and $b_{j}$ are the
moment $j$ of the distribution.%
\begin{equation}
\begin{array}{l}
p^{\ast }=\arg \max -\underset{i}{\sum }p_{i}\log p_{i},\text{ \ s.t.} \\ 
\underset{i}{\sum }p_{i}=1 \\ 
\underset{i}{\sum }h_{j}\left( X_{i}\right) p_{i}=b_{j} \\ 
p_{i}\geq 0\text{ }j=1,...,m\text{, }i=1,...,n.%
\end{array}
\label{Max-entropy}
\end{equation}

The density that respect all the conditions of the model (\ref{Max-entropy})
is defined by \textit{Entropy Density (ED)}. The Lagrangean of the problem is%
\begin{equation}
L=-\underset{i}{\sum }p_{i}\log p_{i}-\lambda _{0}\left[ \underset{i}{\sum }%
p_{i}-1\right] -\sum\limits_{j=1}^{m}\lambda _{j}\left[ \underset{i}{\sum }%
h_{j}\left( X_{i}\right) p_{i}-b_{j}\right] ,  \label{Lagrangean}
\end{equation}%
where $\lambda _{0}$ and $\lambda _{i}$ are the Lagrange multipliers for
each probability or moment constraint. The solution to this problem has the
form 
\begin{equation}
p_{i}=\exp \left[ -\lambda _{0}-1-\sum\limits_{j=1}^{m}\lambda
_{j}h_{j}\left( X_{i}\right) \right] .  \label{solution1}
\end{equation}%
For small values of $m$ it is possible to obtain explicit solutions [Zellner
(1996)]. If $m=0$, meaning that no information is given, one obtains a
uniform distribution. As one adds the first and the second moments, Golan,
Judge and Miller (1996) recall that one obtains the exponential and the
normal density, respectively. The knowledge of the third or higher moments
does not yield to a density in a closed form and only numerical solutions
may provide densities.

In many cases, precise values for moments and probabilities are unavailable.
In face of this problem Abbas (2005) propose the use of the ME principle
using upper and lower bonds in the moments constraints.

There are several research studies of ME in economics. Buchen and Kelly
(1996) describe the application of ME principle to the estimation of the
distribution of an underlying asset from a set of option prices. Samperi
(1999) selects a pricing measure that is consistent with observed market
prices by minimizing the relative entropy functional subject to linear
constraints. This author explores the relationship between entropy, utility
theory and arbitrage pricing theory. Gulko (1998) in its research work
develops the \textit{Entropy Pricing Theory }as a main characteristic of an
efficient market, where the entropy is maximized in order to find the market
equilibrium. Stuzer (1996, 2000) derivates the Black-Scholes model by
solution of a constrained minimization of relative entropy and concludes
that relative entropy minimization provides a simple way to compute the
martingale measure that yields the important Black-Scholes option formula.

The ME principle has been more recently applied in decision analysis,
specially in the specification and estimation of utility values and utility
functions. For example, Fritelli (2000) derives the relative entropy
minimizing martingale measure under incomplete markets and demonstrates the
connection between it and the maximization of exponential utility. Herfert
and La Mura (2004) use a non-parametric approach based on the maximization
of entropy to obtain a model of consumer's preferences using available
evidence, namely surveys and transaction data. In a different approach Abbas
(2004) presents an optimal question-algorithm to elicit von Neumann and
Morgenstein utility values using the ME principle. The same author [Abbas
(2006a)] uses ME to assign utility values when only partial information is
available about the decision maker's preferences and [Abbas (2006b)] uses
the discrete form of ME principle to obtain a joint probability distribution
using lower order assessments. Yang and Qiu (2005) propose an expected
utility-entropy measure of risk in portfolio management, and the authors
conclude that using this approach it is possible to solve a class of
decision problems which cannot be dealt with the expected utility or
mean-variance criterion.

Sandow \emph{et al.} (2006) use the minimization of cross-entropy (or
relative entropy) to estimate the conditional probability distribution of
the default rate as a function of a weighted average bond rating, concluding
that the modeling approach is asymptotically optimal for an expected utility
maximizing investor. Friedman \emph{et al}. (2007) explore an utility-based
approach to some information measures, namely the Kullback-Leibler relative
entropy and entropy using the example of horse races. On the other way,
Darooneh (2006) uses the ME principle to find the utility function and the
risk aversion of agents in a exchange market.

According to Abbas (2006b), the ME principle presents several advantages
when we purpose to construct joint probability distributions and assign
utility values, namely: $\left( i\right) $ it incorporates as much
information as there is available at the time of making the decision; $%
\left( ii\right) $ it makes any assumptions about a particular form or a
joint distribution; $\left( iii\right) $ it applies to both numeric and
nonnumeric variables; and $\left( iv\right) $ it does not limits itself to
the use of only moments and correlation coefficients, which may be difficult
to obtain in decision analysis practice.

\section{Utility and entropy}

When a decision problem is deterministic, the order of the prospects is
enough to define the optimal decision alternative. However, when uncertainty
is present, it is necessary to assign the non Neumann and Morgenstein
utility values. One of the basic assumptions of decision theory is that an
agent's observed behaviour can be rationalized in terms of the underlying
preference ordering and if the observed behaviour is consistent with the
ordering we can infer about the utility function using the available data.
Sometimes the observations are not sufficient to identify clearly the
orderings and one needs more general inference methods. La Mura (2003)
presented a non-paramentric method for preference estimation based on a set
of axiomatic requirements: $\left( i\right) $ no information; $\left(
ii\right) $ uniqueness; $\left( iii\right) $ invariance; $\left( iv\right) $
system independence, and $\left( v\right) $ subset independence. The axioms
characterize a unique inference rule, which amounts to the maximization of
the entropy of the decision-maker's preference ordering.

We extend an approach developed by Abbas (2006a) and also used before in a
similar way by Herfert and La Mura (2004), the \emph{maximum entropy utility
principle}, where a utility function is normalized to range from zero to one
and the utility density function is the first derivative of a normalized
utility function. Based on such definition, the utility density function has
two main properties: $\left( i\right) $ is non-negative; and $\left(
ii\right) $ integrates to unity. The two properties allows the analogy
between utility and probability, and consequently, with entropy [Abbas
(2006a)].

For the discrete case, the utility vector has $K$ elements, defined as%
\begin{equation}
U\triangleq \left( u_{0},u_{1},...,u_{K-2},u_{K-1}\right) =\left(
0,u_{1},...,u_{K-2},1\right) .  \label{vectorU}
\end{equation}%
This vector of dimension $K$ can be represented as a point in a $\left(
K-2\right) $ dimensional space, which is defined by $0\leq u_{1}\leq ...\leq
u_{K-2}\leq 1.$ This region, called utility volume, has a volume equal to $%
1/\left( K-2\right) !.$

In the utility increment vector $\left( \Delta U\right) $ the elements are
equal to the difference between consecutive elements in the utility vector,
it has $K-1$ elements and is defined by%
\begin{equation*}
\Delta U\triangleq \left( u_{1}-0,u_{2}-u_{1},...,1-u_{K-2}\right) =\left(
\Delta u_{1},\Delta u_{2},...,\Delta u_{K-1}\right) .
\end{equation*}%
The coordinates of $\Delta U$ are all non-negative and sum to one.

According to Abbas (2006a) the knowledge of the preference order alone do
not inform at all about the location of the utility increment vector. In
this conditions is reasonable to assume that the respectively location is
uniformly distributed over the domain. The assumptions gives equal
likelihood to all utility values and satisfy agent's preference order,
adding no further information than the knowledge of the order of the
prospects.

For the continuous case, the concepts are similar, but the number of
prospects $K$ can be infinite. Is this case the utility vector is a utility
curve $\left[ U\left( x\right) \right] $, and has the same mathematical
properties as a cumulative probability distribution. The utility increment
vector (or in this case, utility density function) is now a derivative of
the utility curve%
\begin{equation}
u\left( x\right) \triangleq \frac{\partial U\left( x\right) }{\partial x}
\label{utility-density}
\end{equation}%
which is non-negative and integrates to unity.

Given the analogy between utility and probability, the concept of entropy
can be used as a measure of spread for the coordinates of the utility
increment vector%
\begin{equation}
H\left( \Delta u_{1},\Delta u_{2},...,\Delta u_{K-1}\right)
=-\sum\limits_{i=1}^{K-1}\Delta u_{i}\log \Delta u_{i}.
\label{entropy-utility}
\end{equation}
The utility increment vector that maximizes this measure is the uniform
distribution. There are other measures that can be used to spread the
utility increment vector, although, the entropy satisfies the following 3
axioms: $\left( 1\right) $ the measure of spread of the utility increment
vector is a monotonically increasing function of the number of prospects $K$%
, when the utility increments are all equal; $\left( 2\right) $ the measure
of spread of a utility increment vector should be a continuous functions of
the increments; $\left( 3\right) $ the order in which we calculate the
measure of spread should not influence the results.

The differential entropy can also be applied to a utility density function 
\begin{equation*}
H\left( u\left( x\right) \right) =-\int\limits_{a}^{b}u(x)\log u(x)dx,
\end{equation*}%
and this function is maximized when $u\left( x\right) =1/\left( b-a\right) .$
The uniform density integrates to a linear (risk neutral) utility function.

The maximum entropy utility problem is described by%
\begin{equation}
\begin{array}{l}
u_{\max ent}\left( x\right) =-\int\limits_{a}^{b}u(x)\log u(x)dx,\text{ \
s.t.} \\ 
\int\limits_{a}^{b}u(x)dx=1 \\ 
\int\limits_{a}^{b}h_{i}\left( x\right) u(x)dx=b_{i} \\ 
u(x)\geq 0,\text{ }i=1,...,n.%
\end{array}
\label{model}
\end{equation}

Abbas(2006a) used a CARA utility density to show that the differential
entropy has a unique maximum, that occurs exactly when the agent is risk
neutral.

This approach is also defended by Darooneh (2006), that considers that\ the
equilibrium condition may be expressed by the maximum entropy utility, since
the risk of the market induce the randomness. The solution for this problem
is given by the following expression%
\begin{equation}
u_{\max ent}\left( x\right) =\exp \left[ -\lambda _{0}-1-\lambda
_{1}h_{1}\left( x\right) -\lambda _{2}h_{2}\left( x\right) -...-\lambda
_{n}h_{n}\left( x\right) \right] ,  \label{utility-entropy}
\end{equation}%
where $\left[ a,b\right] $ are the domain of the prospects, $h_{i}\left(
x\right) $ is a given preference constraint, $b_{i}^{\prime }s$ are a given
sequence of utility values or moments of the utility function and $\lambda
_{i}$ is the Lagrangean multiplier for each utility value. The uniform
utility density is a special case of equation (\ref{utility-entropy}) where
the constraints $h_{i}\left( x\right) $ do not exist. When $h_{1}\left(
x\right) =x$ and the remaining constraints are zero, the maximum entropy
utility is a CARA utility on the positive domain. When $h_{1}\left( x\right)
=x$ and $h_{2}\left( x\right) =x^{2}$ the maximum entropy utility is a
Gaussian utility density, which integrates to a S-shaped prospect theory
utility function on the real domain.

The risk aversion parameter $\left( \gamma \right) $, using the Arrow-Pratt
definition, of the agent is given by%
\begin{equation}
\gamma _{\max ent}\left( x\right) =-\frac{\partial \ln \left[ u_{\max
ent}\left( x\right) \right] }{\partial x}=\lambda _{1}h_{1}^{^{\prime
}}\left( x\right) +\lambda _{2}h_{2}^{^{\prime }}\left( x\right)
+...+\lambda _{n}h_{n}^{^{\prime }}\left( x\right) ,  \label{risk aversion}
\end{equation}%
where $h_{i}^{^{\prime }}\left( x\right) =\partial h_{i}\left( x\right)
/\partial x.$ The equation (\ref{risk aversion}) shows the linear effect
contributed by the derivative of each preference constraint on the overall
risk aversion function.

Abbas (2006a) presents several examples of application of maximum entropy
utility principle, namely for cases when we know some utility values, cases
when we need to infer utility values by observing decisions and for the case
of multiattribute utility. For all the cases explored, Abbas (2006a)
concludes that the maximum entropy utility principle presents advantages and
satisfies the important assumption of utility and probability independence
that stems from the foundations of normative utility theory.

\section{Conclusions}

This paper presents an efficient alternative way to estimate the utility
function of any agent when there is only partial available information about
the decision maker's preferences. The maximum entropy approach here
presented provides a unique utility function that makes no assumptions about
the structure, unless there is preference information to support it.

Based on the recent literature on this research area, we show that the
analogy probability - utility can be explored in order to use the
information theory measures, and obtain a more robust estimation of the
utility function.

\bigskip

\textbf{Acknowledgement}

Financial support from Fundacao da Ciencia e Tecnologia, Lisbon, is
gratefully acknowledged by the authors, under the contract
PTDC/GES/70529/2006.

\end{document}